\documentstyle[twoside,fleqn,espcrc2,epsfig,amsfonts,latexsym]{article}

\newcommand{\AmS}{{\protect\the\textfont2
  A\kern-.1667em\lower.5ex\hbox{M}\kern-.125emS}}
\newcommand{\bbox}{\lower0.85pt\hbox{$\Box$}}
\newcommand{\kreisl}{\raise0.85pt\hbox{$\scriptstyle\bigcirc$}}
\newcommand{\dreieck}{\raise0.85pt\hbox{$\scriptstyle\bigtriangledown$}}
\newcommand{\stern}{\lower0.85pt\hbox{\Large $\star$}}

\title{ 
\vspace{-8mm}
\rightline{\small UL-NTZ 21/98, KANAZAWA-98-10}
\vspace{-2mm}
\rightline{\small August 20, 1998}
Matter near to the Endpoint of the Electroweak Phase Transition
    }               
 
\author{
E.-M.~Ilgenfritz\address{Institute for Theoretical Physics, Kanazawa University,
Kanazawa 920-1192, Japan},
A.~Schiller\address{Institut f\"ur Theoretische Physik and NTZ, 
Universit\"at Leipzig,
D-04109 Leipzig, Germany}\thanks{Talk at LAT'98 given by A.~Schiller}
and
C.~Strecha$\mathrm{^b}$
}
        
\begin{document}

\begin{abstract}
  Wave functions and the screening mass spectrum in the $3D$ $SU(2)$--Higgs 
  model near to the phase transition line below the endpoint and in the 
  crossover region are calculated. In the crossover region the changing 
  spectrum versus temperature is examined showing the aftermath of the phase
  transition at lower Higgs mass. Large sets of operators with various 
  extensions are used allowing to identify wave functions in position space. 
\end{abstract}

\maketitle

According to recent lattice studies in $3D$~\cite{endpointhkarschendpoint}
and $4D$~\cite{aoki98fodor98} the $SU(2)$--Higgs model ceases to possess a 
first order transition for a Higgs mass $M_H > 72$ GeV. This and the small 
amount of CP violation in the standard model seem to rule out the possibility 
to explain the BAU generation within the standard model.
 
The lattice version of the $SU(2)$--Higgs model is still interesting as a 
laboratory for investigating the behaviour of hot gauge fields coupled to 
scalar matter, for the characterisation of possible bound states, for the 
understanding of real time topological transitions and the role of embedded 
topological defects \cite{CGIS} at the transition. 

The $3D$ lattice model is defined by the action 
\begin{eqnarray}
  S &=&  \beta_G \sum_p \big(1 - \frac{1}{2} \mbox{tr} U_p \big) 
   -  
  \beta_H \sum_{x,\mu} S_{x,\mu}(1)
  \nonumber \\
  & & + \sum_x  \big( \rho_x^2 + \beta_R (\rho_x^2-1)^2 \big) \,.
  \label{eq:S3Dlattice}
\end{eqnarray}
The lattice couplings are related to the continuum parameters of the $3D$ 
$SU(2)$--Higgs model $g_3$, $\lambda_3$ and $m_3$ (see {\it e.g.} 
\cite{wirNP97})
\begin{eqnarray}
  \beta_G &=& \frac{4}{a g_3^2} \,, \
  \beta_H = \frac{2 (1-2\beta_R)}{6+a^2 m_3^2} \,, \nonumber \\
  \beta_R &=& \frac{\lambda_3}{g_3^2} 
  \frac{\beta_H^2}{\beta_G}
  = \frac18 \left( \frac{M_H^*}{80\; \mbox{GeV}}\right)^2 
  \frac{\beta_H^2}{\beta_G}\,,  
  \label{eq:couplings}
\end{eqnarray}
they can be expressed via perturbation theory in terms of $4D$ couplings and 
masses \cite{generic}. The parameter $M_H^*$ is approximately equal to the 
zero temperature physical Higgs mass. The summation in (\ref{eq:S3Dlattice}) is
taken over plaquettes $p$, sites $x$ and links $l=\{x,\mu\}$. The gauge fields 
are represented by unitary $2 \times 2$ link matrices $U_{x,\mu}$, $U_p$ 
denotes the $SU(2)$ plaquette matrix. The Higgs field is parametrised as 
follows: $\Phi_x = \rho_x V_x$ with 
$\rho_x^2= \frac12 \mbox{\rm tr}(\Phi_x^+\Phi_x)$ and $V_x$ is an element of 
the group $SU(2)$, $S_{x,\mu}(1)$ is defined below in (\ref{op:1-}).

Here we report on some results of our recent study \cite{spectrum} of the 
screening spectrum across the very weak phase transition ($M_H^*= 70$ GeV) and 
the crossover region (at $M_H^*= 100$ GeV). This complements earlier studies 
near to a strongly first order transition and at markedly larger Higgs 
mass \cite{philipsen}. Details of the update algorithms have been reported 
before \cite{wirNP97}. 

To study simultaneously the ground state {\sl and} excited states (as well as
their wave functions) one has to consider cross correlations between
(time-slice sums of) operators ${\cal {O}}_{i}$ from a complete set in a given
$J^{PC}$ channel with quantum numbers $J$ (angular momentum) , $P$ (parity)
and $C$ (charge conjugation). According to the transfer matrix formalism, one
should be able to write the connected correlation matrix at time
separation $t$ in the spectral decomposition form
$
  C_{ij}(t) = \sum_{n=1}^{\infty} \Psi_i^{(n)} \Psi_j^{(n)*} e^{-m_n t} 
$
with
$
  \Psi_i^{(n)}=\langle \mathrm{vac}|
  {\cal{O}}_i | {\bf \Psi}^{(n)}\rangle
$,
$|{\bf \Psi}^{(n)}\rangle$ is the $n$-th (zero momentum) energy
eigenstate.  By suitable diagonalisation this allows to find masses
{\it and} wave functions of the lowest mass screening states (ground state)
and higher mass excited states in the various $J^{PC}$ channels.
However, in practice one has to choose a truncated set of operators
${\cal{O}}_i$, ($i=1,\ldots,N$). 

Solving the generalised eigenvalue problem 
$
  C(t)\Psi^{(n)}  =  
  \lambda^{(n)}(t,t_0) C(t_0)\Psi^{(n)}   
$
or
$
  \tilde{C}(t,t_0) \tilde{\Psi}^{(n)}  =  
  {\lambda}^{(n)}(t,t_0) 
  \tilde{\Psi}^{(n)}  \, ,
$  
with
$
  \tilde{C}(t,t_0)= C^{-\frac12}(t_0) C(t) C^{-\frac12}(t_0)
$ 
($t>t_0$, where $t_0=0,1,2$), errors related to this truncation can be kept 
minimal \cite{lueschergattringer}. Practically the decomposition of the matrix 
$C(t_0)$ is performed using a Cholesky decomposition: $C(t_0)=L L^T$. The 
optimised eigenfunctions $\Psi^{(n)}$ in the chosen operator basis (obtained 
with a small distance $t_0$) give an information about the overlap of the 
source operators ${\cal {O}}_i$ with the actual eigenstates 
$|{\bf \Psi}^{(n)}\rangle$. The masses $m^{(n)}$ of these states are obtained 
by fitting the diagonal elements 
$ 
  \mu^{(n)}(t,t_0)= 
  \tilde{\Psi}^{(n)}
  \tilde{C}(t,t_0)
  \tilde{\Psi}^{(n)}$
to a hyperbolic cosine form with $t$ in some plateau region of a local
effective mass.

In contrast to a smearing technique for gauge links and Higgs fields
\cite{philipsen},
we have collected only a few types of operators ${\mathcal{O}}_i$ in our base 
but with a wide span of sizes $l$ in lattice spacings. Such a basis allows to 
obtain information on the spatial extension of a bound state without going 
through a variational procedure. Here, we have restricted ourselves to the 
following operators (with $\mu=3$ reserved for the correlation direction):
\begin{eqnarray}
  0^{++} : \!\!\!\!\! & & \!\!\!\!\! \rho_x^2, \  
  S_{x,1}(l)+S_{x,2}(l), \
  W_{x,1,2}(l)+W_{x,2,1}(l) \nonumber \\
  1^{--} :\!\!\!\!\! & & \!\!\!\!\! V_{x,1}^b(l)+V_{x,2}^b(l),  \  
  2^{++} :   S_{x,1}(l)-S_{x,2}(l)  
  \label{extended}
\end{eqnarray}
where  
\begin{eqnarray}
  S_{x,\mu}(l) \!\!\!\! & = &\!\!\!\!  
  \frac{1}{2}{\mathrm{tr}}(\Phi^+_x U_{x,\mu}\ldots\! 
  U_{x+(l-1)\hat\mu,\mu}\Phi_{x+l \hat \mu}) , \label{op:1-} \\
  V_{x,\mu}^b(l) \!\!\!\! &= &\!\!\!\! 
  \frac{1}{2}{\mathrm{tr}}(\tau^b\Phi^+_x
  U_{x,\mu}\ldots \!
  U_{x+(l-1)\hat \mu,\mu}\Phi_{x+l\hat \mu}) ,  \nonumber
\end{eqnarray}
$W_{x,\mu,\nu}(l)$ are quadratic Wilson loops of size $l \times l $.

Using the cross correlation technique we were able to obtain the wave function
squared corresponding to the optimised operator for each individual state in
the spectrum. Being functions of a physical distance, the squared wave
functions are shown immediately {\it vs.} $l a g_3^2$ in order to overlay
data from measurements at various gauge couplings (lattice spacings) taken
along a line of constant physics near to the transition temperature.

\begin{figure}[!htb]
  \begin{minipage}{7.5cm}
    \begin{center}
      \epsfig{file=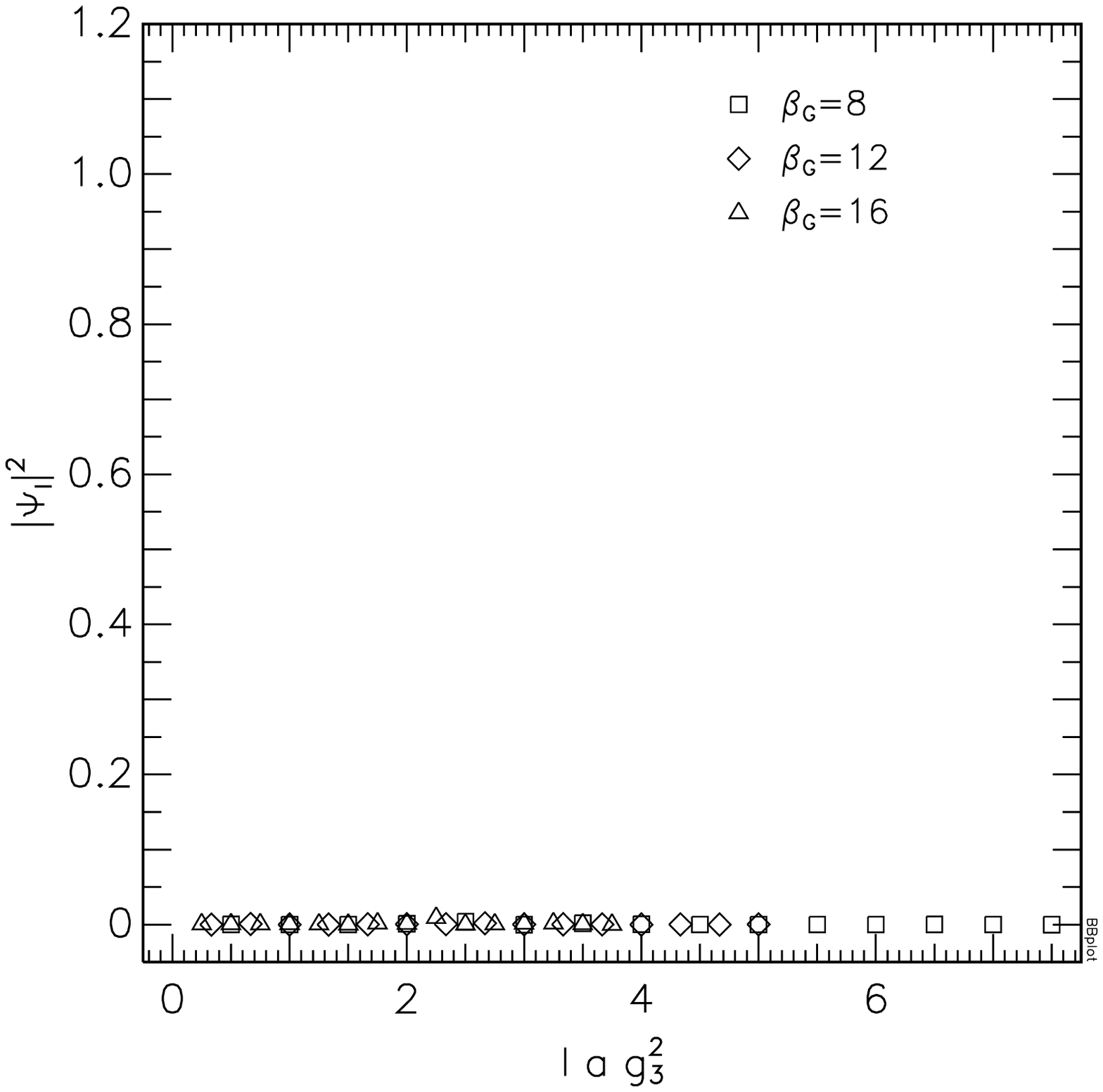,width=3.7cm,height=3.7cm,angle=0}
      \epsfig{file=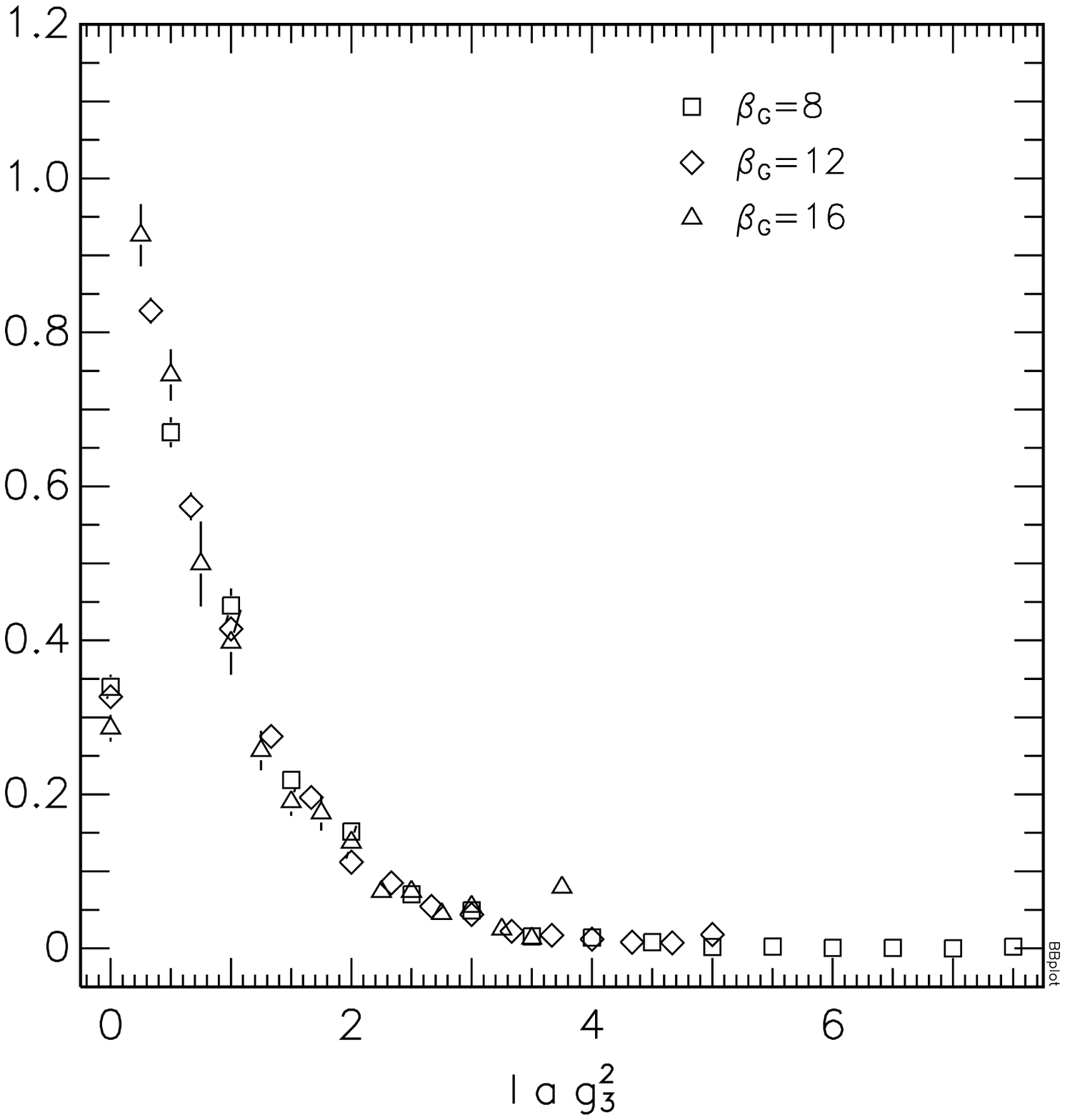,width=3.6cm,height=3.7cm,angle=0}
      \vspace{-15mm}
      \caption[]{\small Squared wave function of the ground state in the 
        $0^{++}$ channel,
        measured on a $30^3$ lattice in the
        symmetric phase; 
        left: $W_{x,1,2}(l)+W_{x,2,1}(l)$,
        right: $S_{x,1}(l)+S_{x,2}(l)$; 
        $l=0,\ldots, 15$}
      \label{fig:kontwave_10+}
    \end{center}
  \end{minipage}
\end{figure}

\begin{figure}[!htb]
\vspace{-9mm}
  \begin{minipage}{7.5cm}
    \begin{center}
      \epsfig{file=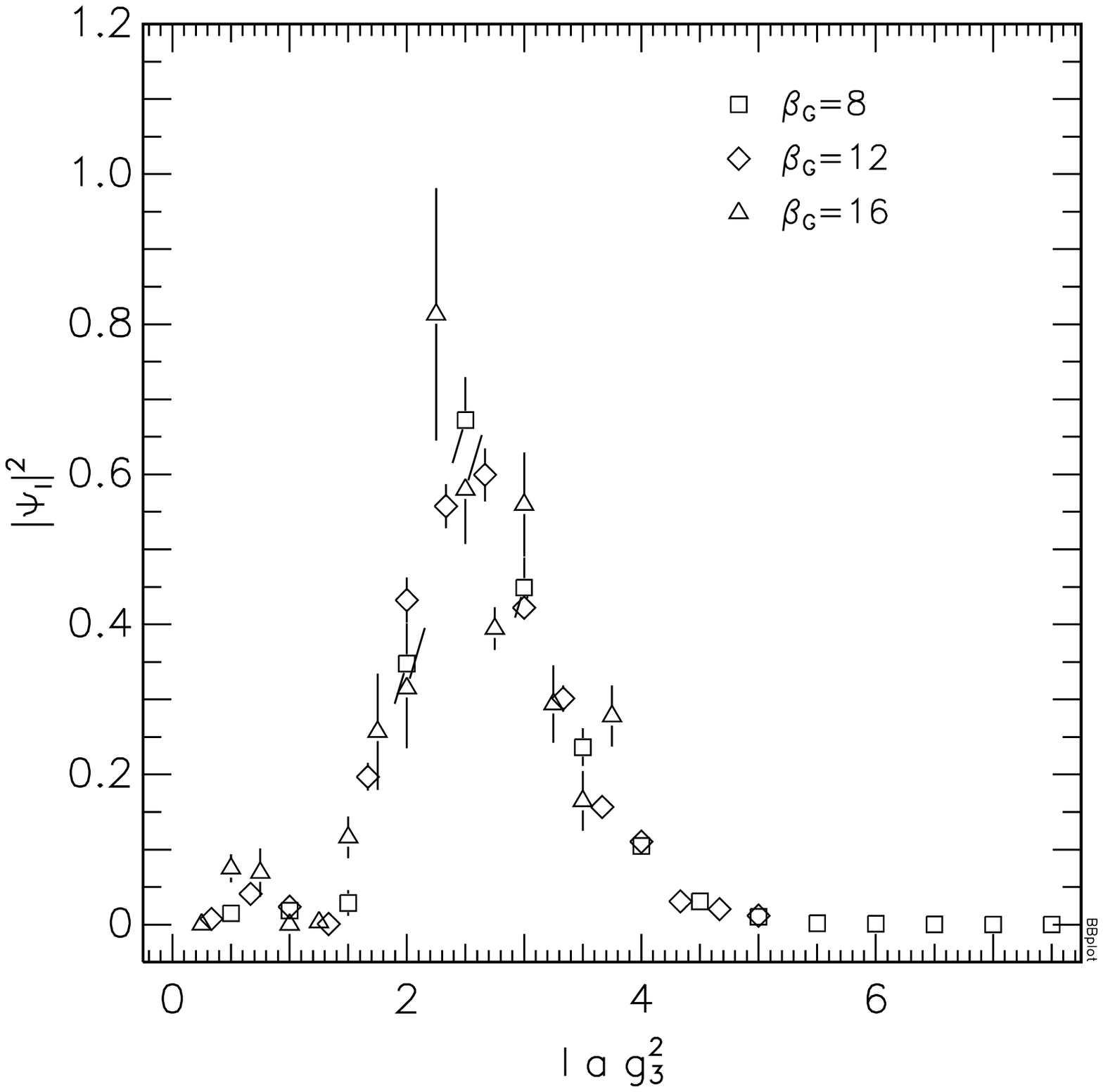,width=3.7cm,height=3.7cm,angle=0}
      \epsfig{file=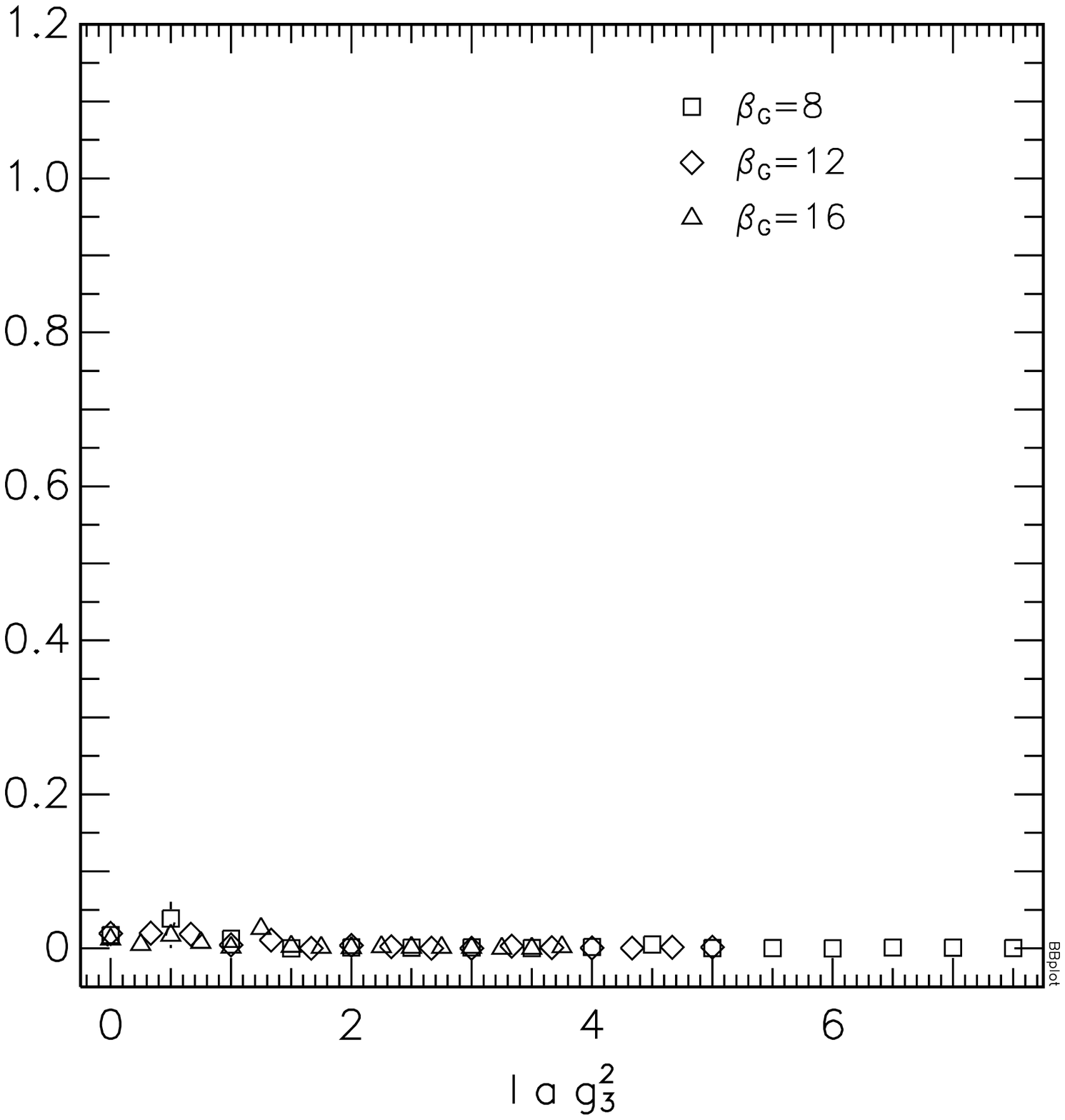,width=3.6cm,height=3.7cm,angle=0}
      \vspace{-15mm}
      \caption[]{\small Same as Fig.~\ref{fig:kontwave_10+} for the
      second excited state}
      \label{fig:kontwave_30+}
    \end{center}
  \end{minipage}
\vspace{-8mm}
\end{figure}

\begin{figure}[!t] 
  \begin{minipage}{7.5cm}
    \begin{center}
      \epsfig{file=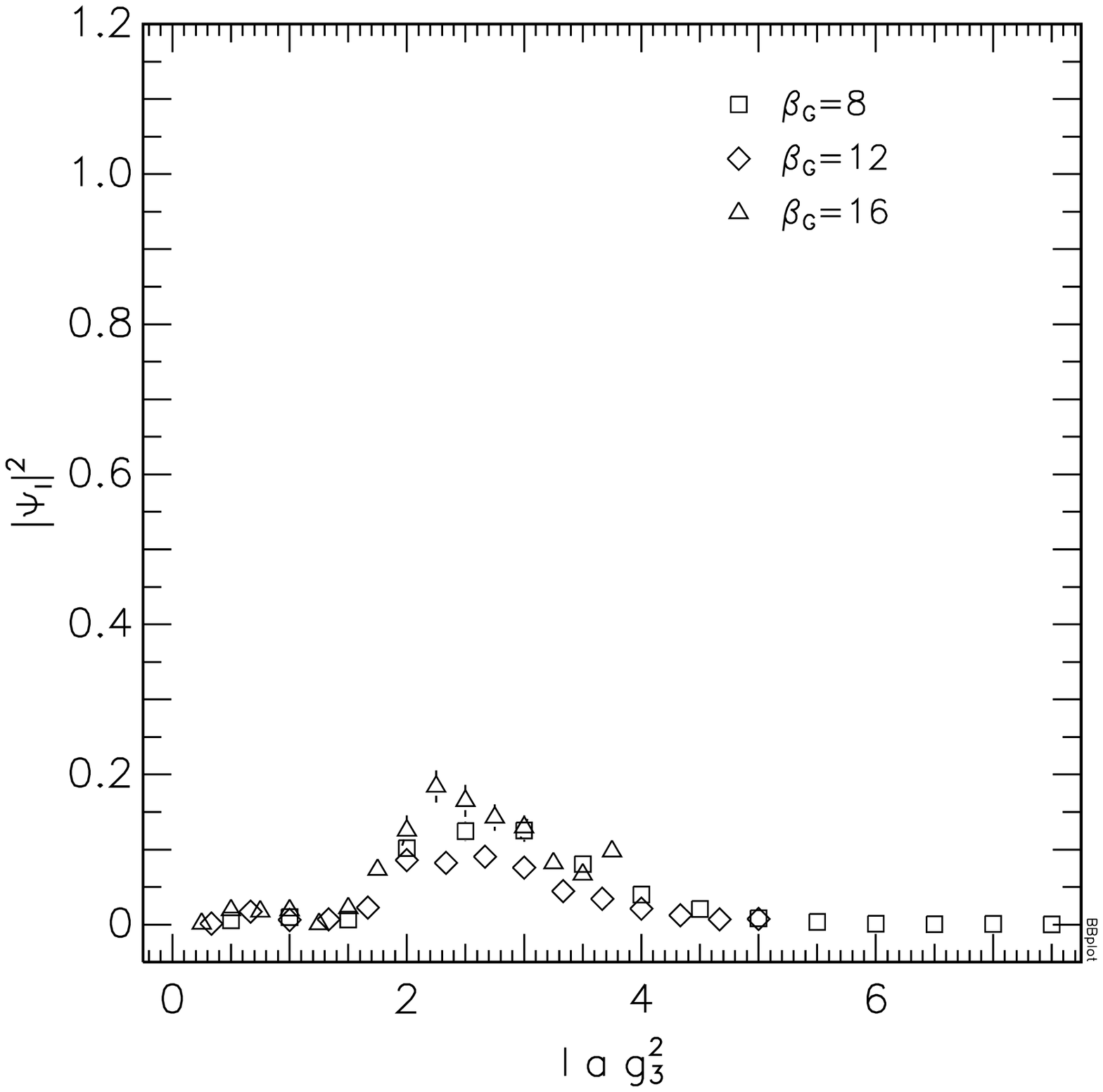,width=3.7cm,height=3.7cm,angle=0}
      \epsfig{file=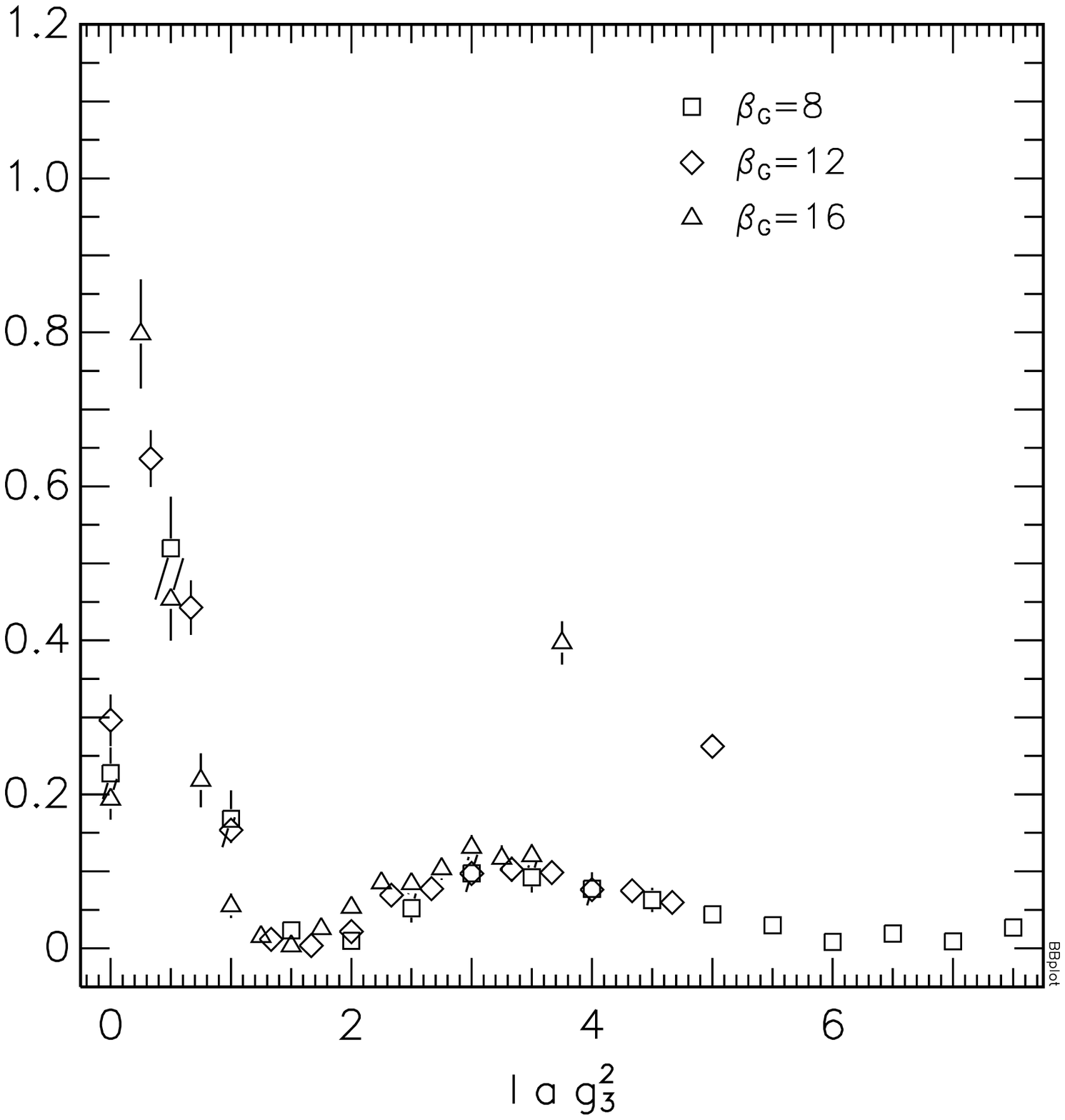,width=3.6cm,height=3.7cm,angle=0}
      \vspace{-15mm}
    \caption[]{\small 
    Same as Fig.~\ref{fig:kontwave_10+} for the first excited state
    in the Higgs phase}
    \label{fig:30b126a_20+} 
    \end{center} 
  \end{minipage}
\vspace{-11mm}  
\end{figure}

Results for the $0^{++}$ channel are collected in
Figs.~\ref{fig:kontwave_10+}-\ref{fig:30b126a_20+} for the squared
\vspace{-8mm}
wave functions. The contributions from the Higgs string and Wilson loop
operators are shown separately in order to identify clearly Higgs and $W$-ball
excitations. In the symmetric phase we observe no mixing of these two operator 
types in the Higgs ground state and the first excitation (not shown). The 
second excited state consists of a pure excitation of gauge degrees of freedom 
(d.o.f.) and can be identified with a $W$-ball in analogy with the glueballs 
of pure $SU(2)$. Our results on this decoupling confirm the observations 
in \cite{philipsen} made at a much lighter Higgs mass.

On the Higgs side of the phase transition pure gauge matter ($W$-ball)
excitations are not expected to be present in the spectrum. In the $0^{++}$ 
channel, our operator set is sufficient to observe a marked difference between 
the phases which is not in accordance to naive expectations. We observe a 
mixing between the two operator types, $W$-ball operators (pure gauge d.o.f.) 
and operators projecting onto Higgs states. This is demonstrated for the first 
excited Higgs state which contains a noticeable contribution from Wilson loop 
operators (Fig.~\ref{fig:30b126a_20+}). 
We interpret this mixing of Higgs and gauge d.o.f. as a signal of the near
endpoint of the phase transition. Deeper in the Higgs phase the contribution
from gauge d.o.f. is expected to disappear which has been checked in our 
simulations at $M_H^*=100$ GeV.
 
The spectrum change has been studied in more detail at $M_H^*=100$ GeV while
{\it continuously} passing the crossover line (changing $\beta_H$) at fixed 
gauge coupling $\beta_G=12$.  We have found a behaviour very similar to our 
results obtained at $M_H^*=70$ GeV. The similarities concern both the high 
temperature side of the crossover (where one expects thermodynamic properties 
being close to those of the symmetric phase at smaller Higgs mass) and the 
region very near to the crossover line on the so-called Higgs side of the 
crossover. 

\begin{figure}[!t]
   \begin{minipage}{7.5cm}
  \centering 
    \epsfig{file=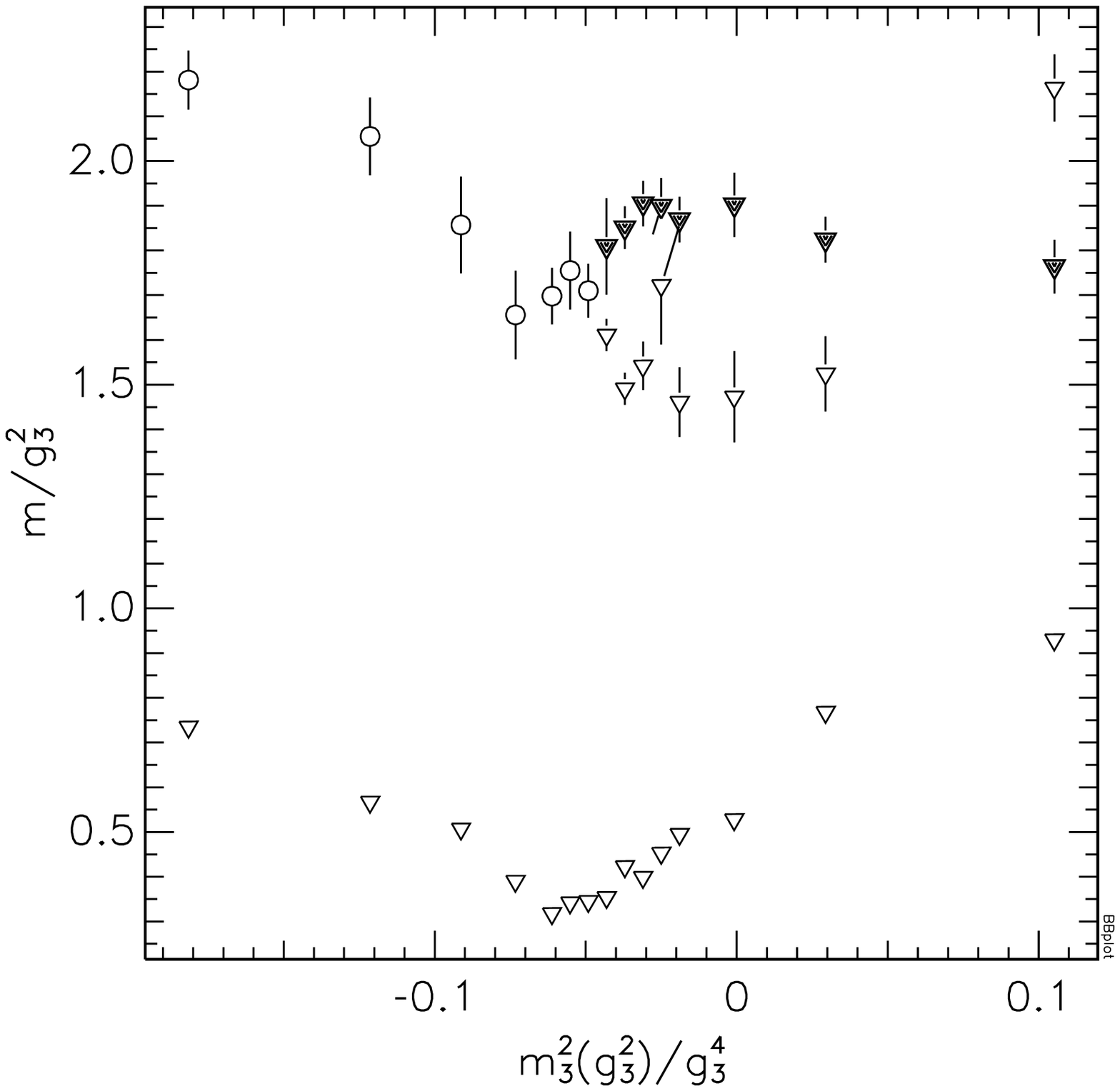,width=3.6cm,height=3.6cm,angle=0} 
    \epsfig{file=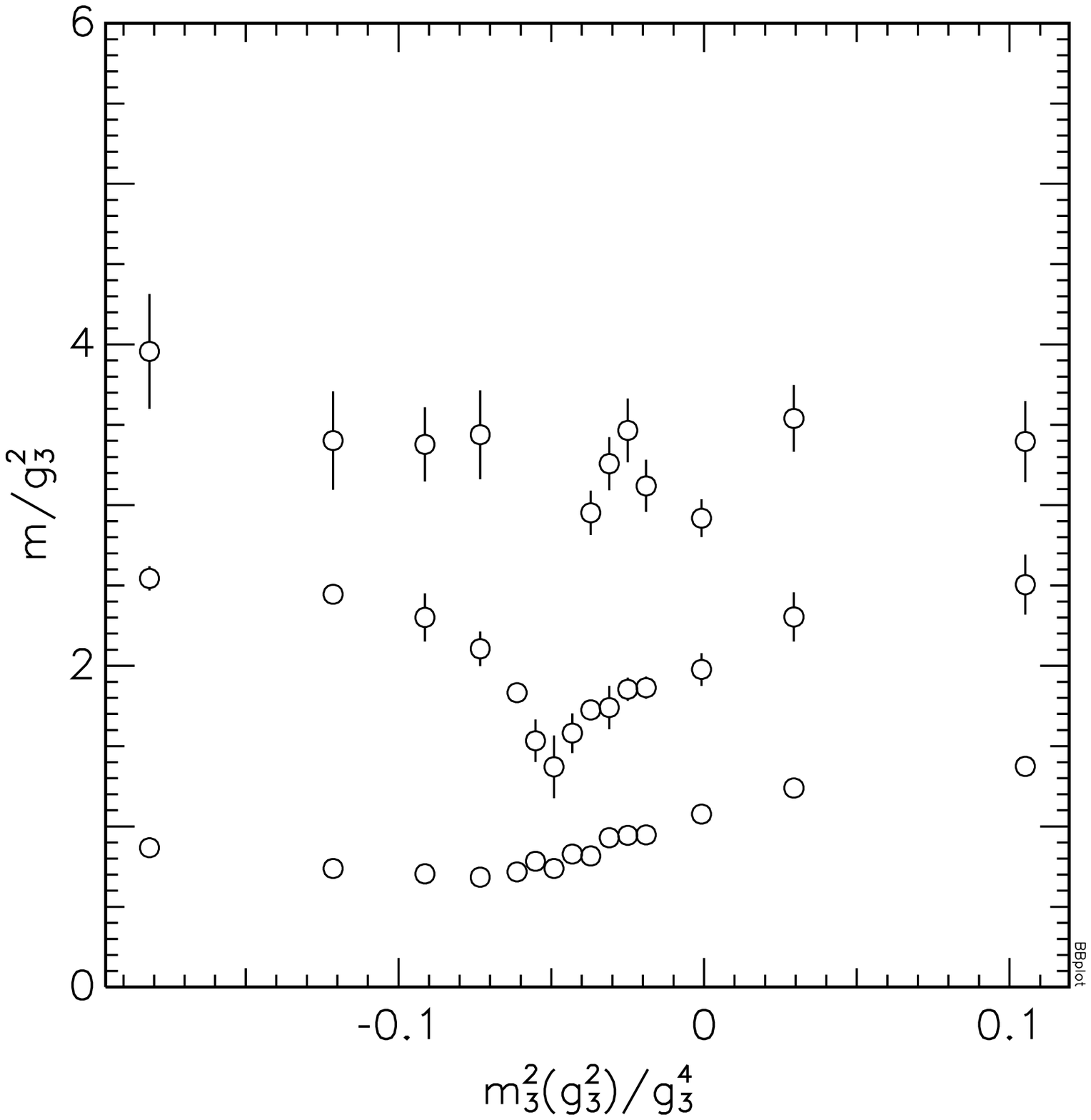,width=3.6cm,height=3.6cm,angle=0}
      \vspace{-10mm}
     \caption[]{\small Screening mass spectra 
     near the crossover in channels $0^{++}$ 
     and $1^{--}$; $0^{++}$: ($\bigtriangledown$)  
     Higgs states, (full
     triangles) $W$-ball states ($\kreisl$) Higgs states 
     with an admixture of
     excited gauge d.o.f.}
    \label{fig:mphi_0++1--}
  \end{minipage}
\vspace{-10mm}
\end{figure}

In Fig.~\ref{fig:mphi_0++1--}
we present the spectrum of the lowest states in the $0^{++}$ and  $1^{--}$ 
channels as function of $m_3^2(g_3^2)/g_3^4$ (or $\beta_H$) over a
certain interval above and below the crossover ({\it i.e.} in temperature).
Looking at the excited states in the $0^{++}$ channel of 
Figs.~\ref{fig:mphi_0++1--} we conclude that the scalar and gauge sector are 
approximately decoupled as long as one keeps away from the crossover line on 
the high temperature side. The mass of the lowest $W$-ball state (full 
triangle) is roughly independent of $\beta_H$ as long as one does not come too 
close to the crossover. Thus, on the high temperature side of the crossover, 
the ordering of states qualitatively resembles the spectrum at smaller values 
of Higgs self-coupling (at $M_H^*=70$ GeV). 

If one approaches the crossover temperature the mass of the first (Higgs-like) 
excitation is moving up towards the lowest $W$-ball state whose mass decreases.
At some point we observe a growing admixture to the Higgs excitation by 
contributions from Wilson loop operators. We have explicitly checked that at 
higher $\beta_H$ (deeper in the would-be Higgs phase) the admixture from pure 
gauge d.o.f. disappears again from this state.  

To summarise, our operator choice was the simplest one to incorporate the 
notion of spatial extension. The continuum limit of the screening masses and 
wave functions has to be accompanied by correspondingly larger lattices with 
the same physical volume, without automatically enlarged operator basis. 
Therefore, in a next step of improvement of the method, smearing of the fields 
entering our operators and/or the construction of blocked operators will become
necessary for further optimising the resolving capability of the method for 
excited states.

\end{document}